\documentclass{PoS}

\title{Constraints on cosmological parameters}

\ShortTitle{Constraints on cosmological parameters}

\author{\speaker{Amedeo Balbi}\\%
        Dipartimento di Fisica \& INFN\\
        Universit\`a di Roma Tor Vergata\\
        Via della Ricerca Scientifica, 1\\
        I-00133, Roma, Italy\\
        E-mail: \email{balbi@roma2.infn.it}}

\abstract{A cosmological model with total density close to critical (and flat geometry), dominated by dark matter and dark energy of unknown nature, and consistent with the basic predictions of the inflationary scenario is a very good fit to a variety of cosmological probes: the anisotropy of the CMB, the large scale distribution of matter, the luminosity distance of high-redshift type Ia supernovae and so on. These high-quality data have established a new standard of precision in the determination of cosmological parameters.}

\FullConference{CMB and Physics of the Early universe\\
		 20-22 April 2006\\
		 Ischia, Italy}

\begin{document}

\section{The standard cosmological model}

Cosmology has a standard model, which provides a well-established
framework to understand the global properties of the physical
universe. There is a strong interplay between fundamental physics and
cosmology, since the early universe is a natural laboratory for high
energy physics. 

The big bang model (or, more precisely, the Friedmann-Robertson-Walker
model) provides a very successful description of the physical universe
from very early times ($t\sim 10^{-2}$ s) to the present. It can
easily explain some key features of the observed universe, such as:
\begin{itemize}
\item the expansion law
\item the abundance of light elements
\item the existence of the cosmic microwave background
\item the age of the oldest objects observed
\end{itemize}
Furthermore, it provides a framework where the gravitational
instability scenario that explains the growth of cosmic structures can
be easily accommodated. 

An additional ingredient of the standard cosmological model, needed to extend the description of the universe back to very early times ($t\sim 10^{-43}$ s after the big bang) is a period of {\em cosmological inflation}, i.e.\ a very short time ($t\sim 10^{-35}$ s) during which the universe expanded exponentially (changing its size of a factor $\sim 10^{30}$). The inflation mechanism has proved quite powerful as a
refinement of the classic big bang model, and is now considered an
important element of the standard cosmological model. Independently
of the details of the specific model, the inflationary scenario makes
a number of testable predictions:
\begin{itemize}
\item The universe must be very close to flat
\item Primordial density perturbations in the universe are gaussian
  distributed, adiabatic and have a power-law power spectrum
\item A stochastic background of gravitational waves should be present
  in the universe
\end{itemize}
Although  no universally accepted and tested model for inflation currently exists, there are a number of viable candidates, all of which are based in one way or
another on the dynamics of a weakly coupled, homogeneous scalar field
$\phi$ \cite{Linde 1982, Albrecht & Steinhardt 1982}. In its simplest
form, the equation of motion of such a field is:
\begin{equation}
\ddot\phi + 3H\dot\phi + V'(\phi)=0
\end{equation}
and its energy density and pressure are given by:
\begin{eqnarray}
\rho_\phi&=&{1\over 2}\dot\phi^2+V(\phi)\\ 
p_\phi&=&{1\over 2}\dot\phi^2-V(\phi)
\end{eqnarray}
Here $H$ is the expansion rate of the universe, and $V$ is the scalar field potential. A common
solution to the field equation of motion is based on the so-called
{\em slow-roll} approximation, which assumes that the field
acceleration $\ddot\phi$ is negligible, so that :
\begin{equation}
3H\dot\phi \simeq- V'(\phi)
\end{equation}
The conditions for the slow-roll assumption to hold are given by:
\begin{equation}
\epsilon \equiv {1\over 16\pi G}\left({V'\over V}\right)^2 \ll 1; \;\;\;\;\;\;
\vert\eta\vert\equiv {1\over 8\pi G}\left\vert{V''\over V}\right\vert\ll 1
\end{equation}
where $\epsilon$ and $\eta$ are called {\em slow-roll parameters}.
Constraining the slow-roll parameters by measuring the
exact shape of the power spectrum of primordial perturbations can rule
out specific models of inflation.

One important feature of inflation is that it provides a mechanism to
generate super-horizon primordial density perturbations in the early
universe.  Broadly speaking, the mechanism goes as follows: consider a
generic quantum fluctuation $\delta\phi(\vec{x},t)$ in the scalar
field $\phi$. The Fourier expansion coefficients of this fluctuation
are $\delta\phi_k$. During inflation the wavelength of each Fourier
component will rapidly grow much bigger than the causal horizon. When
this happens, the corresponding fluctuation will ``freeze'', since no
causal mechanism will be able to influence its evolution. At later
times, long after inflation ends, each wavelength will re-enter the
horizon, and the associated component of the fluctuation will be seen
as a density perturbation. Note that there is no way of producing such
a mechanism in classical cosmology: in the standard big bang model, a
certain comoving scale becomes smaller than the causal horizon at some
given time, and remains inside the horizon ever after.  In a similar
way, inflation also produces a stochastic background of gravitational
waves. Gravitational waves correspond to tensor perturbations in the
space-time metric, while density perturbations are scalar. Density
perturbations produced during inflation are {\em adiabatic}, or {\em
  isentropic}: they are genuine curvature perturbations in the
spacetime metric and leave the ratio of matter and radiation (or of
any other two species) constant at any point in space. Furthermore,
they are {\em gaussian distributed} (or very close to gaussian). The
power spectrum of density perturbations produced by inflation in the
slow-roll approximation is quite simple:
\begin{equation}
P_s(k)=A_s k^{n_s}; \;\;\;\;\;\; P_t(k)=A_t k^{n_t}
\end{equation} 
for scalar and tensor density perturbations respectively, with:
\begin{equation}
n_s=1-4\epsilon+2\eta; \;\;\;\;\;\; n_t=2\epsilon
\end{equation}
Of course, since in the slow-roll regime $\eta$ and $\epsilon$ must
both be very small, inflationary models usually predict a scalar
spectral index very close to 1, a property termed {\em
  scale-invariance}. Similarly, the power spectrum of tensor
perturbations should be roughly constant, since $n_t\simeq 0$. The
ratio of the amplitude of tensor and scalar perturbations must satisfy
the so-called consistency relation $r\equiv A_t/A_s = 13.6\epsilon$.
Measuring the power spectrum of density perturbations is then a
powerful tool to test the inflationary parameters.

\section{The cosmological parameters}

The evolution of the universe in the big bang model is essentially
determined by its content. The density of each component $i$ is measured by its value $\Omega_i$ in units of a critical value ($\rho_c=3H^2/8\pi G$), so that $\rho_i\propto\Omega_i H^2$. The total density parameter in a
multi-component universe is the sum of the density parameters of the
single components:
\begin{equation}
\Omega=\sum_i \Omega_i.
\end{equation}
Assuming that each component has an equation of state of the form
$p=w\rho$, with $w$ independent of time, the Friedmann equation describing the evolution of the universe can be
written as:
\begin{equation}
\left({\dot{a}\over a}\right)^2=H_0^2\left[\sum_i 
\Omega_i a^{-3(1+w_i)}+
(1-\Omega) a^{-2}\right]
\end{equation}
where $a$ is the scale factor parameterizing the expansion, and the density parameters are evaluated at present time. One of the
main tasks of observational cosmology is to obtain accurate estimates
of the parameters in the right hand side of the Friedmann equation:
the Hubble constant $H_0$ and the contributions to $\Omega$ from the various components in the universe. In addition to this, one needs to get some estimate of the parameters defining the inflationary model, such as the amplitude and spectral index of the primordial power spectrum (for both scalar and tensor perturbations). Let us review the status of our knowledge on these parameters.

\subsection{Hubble constant}

The present expansion rate of the universe is measured by the Hubble constant, often parameterized in terms of the adimensional quantity $h$ as
$H_0=100~h$ Km s$^{-1}$Mpc$^{-1}$.  The best value of the Hubble constant comes from the measurements of the Hubble Space Telescope Key Project \cite{Freedman et al. 2001}, which calibrated the cosmic distance scale by observing Cepheids in nearby and distant galaxies. This resulted in a value $H_0=72\pm 3{\rm(statystical)}\pm 7 {\rm (systematical)}$ km/s/Mpc. The value derived from WMAP 3-year observations of CMB anisotropy under the assumption that the universe is flat is in remarkable agreement with this: $H_0=73.4^ {+2.8}_ {-3.8}$ km/s/Mpc \cite{Spergel et al. 2006}.

The present age of the universe is directly related to the Hubble parameter by an integral over the redshift $z$, as $t_0=\int_0^{\infty} dz /(1+z)H(z)$. A lower 2$\sigma$ limit of $11.2$ Gyr comes from observations of globular clusters in the Milky Way (\cite{Krauss & Chaboyer 2003} and refs.\ therein), while WMAP can only constrain the present age when a flat universe is assumed. Under this assumption, $t_0=13.73^{+0.13}_{-0.17}$ Gyr \cite{Spergel et al. 2006}.

\subsection{Total density}

The total density of the universe $\Omega$ also specifies its spatial curvature: the space-time metric has flat spatial sections when $\Omega=1$. 
The best way to determine $\Omega$ is through the position of acoustic features in the angular power spectrum of CMB temperature fluctuations. This provides a direct measurement of the angular size of sound horizon at recombination ($\sim 300\,000$ years after the big bang), which is strongly dependent on $\Omega$. Strong evidence that the universe is flat (as predicted by inflation) first came from balloon-borne experiments such as MAXIMA and Boomerang \cite{Balbi et al. 2000, de Bernardis et al. 2000}. WMAP confirmed this results to greater precision: deviations of $\Omega$  from unity are currently constrained to $-0.015^{+0.020}_{-0.016}$ and will improve of at least an order of magnitude when Planck will obtain its data.

\subsection{Radiation density} The radiation component of the 
universe (relativistic particles) has equation of state
$p_R=\rho_R/3$. When the universe is radiation dominated, the scale
factor evolves as $a\propto t^{1/2}$. According to the standard
cosmological model, today the radiation in the universe is made of the
cosmic microwave background photons and 3 species of relic massless
neutrinos. The present radiation density can be expressed in terms of
the photon temperature $T$, as:
\begin{equation}
\rho_R={\pi^2\over 30} g_\star T^4
\end{equation}
where $g_\star$ counts the total number of effectively massless
degrees of freedom. This can be computed, giving $g_\star=3.36$, while
the cosmic microwave background average temperature is accurately
measured to be $T=2.725\pm 0.001$~K \cite{Fixsen et al. 1996}. Thus, today the radiation gives a
totally negligible contribution to the critical density: $\Omega_R
=4.31\times 10^{-5} h^{-2}$.

\subsection{Matter density}
The equation of state of matter, or non-relativistic particles, is
$p_M=0$, so that during matter domination the scale factor evolves as
$a\propto t^{2/3}$. The most familiar contribution to matter in the
universe comes from baryons (or nucleons). The abundance of light
elements produced in the early universe is strongly dependent on the
baryon-to-photon ratio, which is directly related to the present
baryon density. Measurement of primordial abundances of D, $^3$He,
$^4$He, $^7$Li are a strong probe of the baryon density, and indicate
that baryons contribute to roughly 5$\%$ of the critical density \cite{Fields & Sarkar 2006}. A consistent result is obtained from CMB anisotropy observations, which provide a tight constraint on the baryon density since the ratio of acoustic peak heights is strongly dependent on the baryon-to-photon ratio in the universe. If $\Omega\sim 1$, as predicted by inflation and now accurately confirmed
by cosmological observations, most of the universe is not made of the
same stuff we are made of.

There is strong observational evidence that a large contribution
(about 30$\%$) to the critical density comes from so-called {\em dark
  matter}. This constraint comes from a variety of large scale structure probes, such as the shape of matter power spectrum measured from galaxy redshift surveys such as SDSS or 2dF, the gas fraction in clusters of galaxies, the study of peculiar velocity fields, and so on. Similar results come from the height of the third acoustic peak in the CMB power spectrum observed by WMAP, although better data would be needed to obtain a robust estimate. Theoretically, the most plausible candidate for dark matter
is some heavy, weakly-interacting massive particle, left from the very
early stages of the evolution of the universe. The standard picture
for the production of such a relic is as follows. The candidate
particle is assumed to be initially in thermal equilibrium with the
primordial plasma, so that its abundance decreases as $\exp{(-M_X/T)}$
where $M_X$ is the particle mass and $T$ is the photon temperature.
When the interaction rate of the particle, $\Gamma$, becomes smaller
than the expansion rate of the universe, $H$, the particle decouples
from the thermal plasma and its abundance becomes constant (a moment
known as {\em freeze-out}). Then, a cosmologically relevant relic
abundance can be achieved provided the particle has a large enough
mass, and a small enough interaction rate. There are many candidates
for dark matter (for example, supersymmetric partners): unfortunately,
since it interacts so weakly, direct detection of dark matter proves
challenging.  Some light on the nature of dark matter can be shed by
accurate measurements of its present density by cosmological
observations. 

\subsection{Dark energy}

In its most general form, Einstein equation includes a so-called
cosmological term $\Lambda$ in addition to the familiar stress-energy
tensor:
\begin{equation}
R_{\mu\nu}-{1\over 2}g_{\mu\nu} R=8\pi G T_{\mu\nu}+\Lambda g_{\mu\nu}
\end{equation}
Adding a cosmological constant term is completely equivalent to
introducing a new contribution to the stress-energy tensor from a
component with:
\begin{equation}
\rho_V=\Lambda/8\pi G; \;\;\;\;\;\; p_V=-\Lambda/8\pi G
\end{equation}
It can be shown that this is exactly the kind of contribution
resulting from zero-point fluctuations of quantum fields, or {\em
  vacuum energy}. The equation of state of vacuum energy is
$p_V=-\rho_V$, and the universe expands exponentially when it is
vacuum dominated: $a\propto \exp{\left[(\Lambda/3)^{1/2} t\right]}$.

The evidence in favour of a non-null cosmological constant has been mounting over the past few years. First, measurement of the luminosity distance of high-redshift type Ia supernovae (a particularly good kind of standard candles) can only be explained by a recent accelerated expansion of the universe, a behaviour that cannot be obtained within a Friedmann model containing only matter. Moreover, as we just mentioned, the universe seems to be very close to flat (i.e.\ it has a total density equal to the critical value) but there is not enough matter (either baryonic or non baryonic) to explain the observed flatness. The observed accelerated expansion and the need for a substantial contribution to the cosmic budget, in addition to other clues, such as the need to reconcile the age of the universe to that of the oldest globular clusters, point toward the existence of a cosmologically significant amount of vacuum energy: $\Omega_\Lambda\sim 0.7$.

Unfortunately, the introduction of this seemingly harmless contribution to the energy density of the universe has disturbing implications.
First of all, any estimate of plausible values for the vacuum energy
density from fundamental physics exceeds the critical density $\rho_C$
by at least 40 (and often as much as 120) orders of magnitude, while observational cosmology sets the total energy density of the universe at roughly the critical value, $\Omega\sim 1$. One might hope that some mechanism is leading to an exact cancellation of the contributions to the vacuum energy, so that it is exactly $\rho_V=0$: however, such a mechanism is currently unknown. The situation is even more puzzling, since (as we just mentioned) recent observations of distant type Ia supernovae \cite{Perlmutter et al.  1999, Riess et al. 2001} have shown that we live in a universe that
has just entered a vacuum dominated epoch, starting a phase of
accelerated expansion. This means that the cosmological constant term
is still very small compared to theoretical estimates, but it is large
enough ($\rho_V/\rho_c\sim 0.7$) to be cosmologically relevant in the
present universe. There seems to be a serious fine-tuning problem: if
$\Lambda$ is non-zero, then why is it so small? Furthermore, given the
observed value of $\Lambda$, vacuum-energy was never important in the
past evolution of the universe, but it is starting to be the dominant
contribution at present time. We then seem to live in a very special
moment in the universe: an annoying coincidence indeed.

The vacuum energy problem may in fact be the biggest mystery of modern
physics \cite{vacuum}. A possible way to alleviate it, and one that
has interesting and testable implications for cosmology, is to
consider a generalization of the cosmological constant term, that has
been termed {\em dark energy}. As shown when discussing inflation, a
scalar field $\phi$ with effective potential $V(\phi)$ has an equation
of state with $w=(\dot\phi^2/2-V)/(\dot\phi^2/2+V)$. Any value of $w$
such that $1+3w<0$ results in an accelerated expansion, so it is
dynamically equivalent to a cosmological constant. The interesting
feature of these models is that they admit {\em tracking solutions},
in which the dark energy can reach the present value starting from a
very different set of initial conditions. This mitigates the fine
tuning and coincidence problems but, of course, leaves open the
questions about the nature of the field $\phi$. Cosmological
constraints to $w$ can be able to discriminate among dark energy
models by saying something about the scalar field potential $V$.  The phenomenology of dark energy can in principle be described in terms of a small number of parameters. In addition to the equation of state $w$ (which in general is a function of cosmic time) an important quantity is the sound speed $c_s^2\equiv \delta p/\delta \rho$, which is crucial to describe the clustering behaviour of dark energy, if any. The sound speed needs not be the usual adiabatic one, but can be modified to account for entropy fluctuations, thus encompassing a broad set of dark energy candidates.
Other explanations of the puzzle could lie in the gravity sector, involving some modifications of the left-hand side of Einstein's equation (e.g.\ the existence of non-minimal couplings, or higher dimensional theories, or the feedback from inhomogeneities). An
excellent review on dark energy from the point of view of both
cosmology and fundamental physics is \cite{padma}.

But how to constrain the properties of dark energy? The crucial observable is the evolution of the Hubble parameter, giving a record of the expansion history of the universe. This can be probed in several ways: through the measurement of the luminosity distance of high redshift supernovae, through the determination of the angular diameter distance from the CMB angular power spectrum or from baryonic acoustic features in galaxy redshift surveys,  through the age of the universe, and so on. These observations are all related to the background cosmology, but clustering can play an important role in discriminating dark energy candidates. In fact, theoretical prediction can be grossly misestimated when the behaviour of dark energy perturbations is not properly taken into account. The main observables that can be used to track the effects of clustering are the evolution of the number of sources in the comoving volume, as estimated for example from galaxy cluster counts obtained using the Sunyaev-Zeldovich effect, or the measurement of shear convergence from weak lensing. 

A very promising tool is the cross-correlation of CMB temperature anisotropy maps to tracers of the local distribution of dark matter, such as the projected galaxy distribution. This allows to extract the signal from the integrated Sachs-Wolfe (ISW) effect \cite{Sachs & Wolfe 1967}, which is due to the evolution of gravitational potential along the line of sight. The ISW is strongly affected by dark energy and is a powerful way of discriminating theoretical models. It is hard to detect the ISW in CMB maps because of the unavoidable cosmic variance which is dominant on large angular scales, exactly where the ISW signal is expected to peak. However, the cross-correlation of CMB maps and large scale structure probes can enhance the detection power and lead to significant constraints. The first detection of the ISW \cite{Boughn & Crittenden 2004, Boughn & Crittenden 2005} was obtained  by combining the WMAP 1st year CMB data with the hard X-ray background observed by the High Energy Astronomy Observatory-1 satellite (HEAO-1 \cite{Boldt 1987}) and with the radio galaxies of the NRAO VLA Sky Survey (NVSS \cite{Condon et al. 1998}). The positive correlation with NVSS was later confirmed by other authores, including the WMAP team \cite{Nolta et al. 2004}. Other large scale structure tracers that led to similar positive results were the APM galaxy survey \cite{Maddox et al. 1990}, the Sloan Digital Sky Survey (SDSS \cite{York et al. 2000}) and the near infrared 2 Micron All Sky Survey eXtendend Source Catalog (2MASS XSC \cite{Jarrett et al. 2000}) \cite{Fosalba et al. 2003, Scranton et al. 2003, Fosalba & Gaztanaga 2004, Afshordi et al. 2004, Padmanabhan et al. 2005, Cabre et al. 2006, Vielva et al. 2006}. All of these results collected additional evidence of the existence of a dark energy component (or something with the same gravitational behaviour) which is currently dominating the cosmic expansion. However, much work remains to be done to extract further information on the detailed nature of dark energy: at the moment, there are no strong indications that $w$ is different from the cosmological constant value $-1$ or that it significantly evolved over time.  Better data are needed, especially from future and ongoing redshift surveys.

\subsection{Inflationary parameters}

The recent 3-year WMAP data significantly constrained the space of inflationary parameters. The Harrison-Zeldovich primordial spectrum, with $n_s=1$ and a null ratio of tensor to scalar perturbation ($r=0$), seems to be disfavoured by the data, although it is still within the $95\%$ confidence level \cite{Kinney et al. 2006}. The WMAP results seem also to show a preference for a $m^2\phi^2$ potential over a $\lambda \phi^4$. A significant deviation from a unit spectral index is also found when a combination of data from the Lyman-$\alpha$ forest, galaxy clusters, supernovae and CMB is considered \cite{Seljak et al. 2006}. The WMAP upper limits on tensor perturbations are $r<0.55$ at 2$\sigma$, while this gets tighter ($r<0.28$) when SDSS data are included in the analysis. However, this bounds get looser when the scalar spectral index is allowed to vary with $k$, as predicted by some inflationary models. With its extended lever arm in multipole space, Planck will be able to significantly strenghten the constraints on inflation.

\subsection{Reionization}

The hydrogen in the universe is completely reionized at redshift at least as high as $5$. The exploration of the so called {\em dark ages}, i.e.\ the time before the formation of the first structures in the universe is an active subject of investigation \cite{Choudhury & Ferrara 2006}. The CMB temperature anisotropy signal gets damped when the photons are diffused by free electrons along the line of sight. The amount of damping is a powerful probe of the optical depth to reionization. Furthermore, CMB polarization is generated when the photons last scatter on the free electrons: if the optical depth is non zero, a recognizable polarization signature gets generated at large angular scales, allowing to investigate the detailed reionization history, discriminating models that have the same optical depth but a different evolution of the ionization fraction with redshift \cite{Hu & Holder 2003}. It is not possible to get into this kind of details with present data, but sheding some light on the dark ages should be well within the capabilities of Planck. WMAP data currently constrain the optical depth at a value of roughly $\tau\sim 0.1$, consistent with complete reionization at $z\sim 10$. 

\section{Conclusions}
Cosmology has developed into a fully mature science. The parameters of
the big bang model are now known with great accuracy, and the
constraints are expected to get tighter in the future. Inflation has
not been falsified, and its main predictions are strikingly consistent
with observations. The results obtained using completely different
cosmological probes are in remarkable agreement among themselves, as
well as with theoretical predictions.  Nonetheless, many fundamental
questions are still open \cite{open questions}. The pace of
experimental and theoretical progress, however, does not seem to be
close to a halt.


\begin{thebibliography}{99}
\bibitem{Linde 1982}  A.~D.~Linde, 1982, {\em Phys. Lett.} 108B, 389
\bibitem{Albrecht & Steinhardt 1982} A.~Albrecht  \& P.~J.~Steinhardt,
  1982, {\em Phys. Rev. Lett.} 48, 1220
  \bibitem{Freedman et al. 2001}  W.~L.~Freedman et al., 2001, {\em ApJ}, 553,  47
\bibitem{Spergel et al. 2003} D.~N.~Spergel et al., 2003, {\em ApJS},  148, 175 
\bibitem{Spergel et al. 2006} D.~N~Spergel et al., 2006 [arXiv:astro-ph/0603449]
\bibitem{Krauss & Chaboyer 2003} L.~M.~Krauss \& B.~Chaboyer 2003 {\em Science} 299, 65
\bibitem{Balbi et al. 2000}  A.~Balbi et al. 2000, {\em ApJ}, 545, L1
\bibitem{de Bernardis et al. 2000} P.~de Bernardis et al., 2000,
  {\em Nature}, 404, 955
\bibitem{Fixsen et al. 1996} 
  D.~J.~Fixsen et al. 1996, {\em ApJ}, 473, 576
\bibitem{Fields & Sarkar 2006} B.~Fields \& S.~Sarkar 2006, astro-ph/0601514
\bibitem{Perlmutter et al. 1999} S.~Perlmutter et al., 1999, {\em ApJ}, 517,
  565
\bibitem{Riess et al. 2001} A.~G.~Riess et al., 2001, {\em ApJ}, 560, 49
 \bibitem{vacuum} S.~E~Rugh \& H.~Zinkernagel , 2002, {\em Studies in History and Philosophy of Modern Physics}, 33, 663
\bibitem{padma} T.~Padmanabhan, 2003, {\em Phys. Rept.} 380, 235
\bibitem{Sachs & Wolfe 1967} R.~K.~Sachs\& A.~M.~Wolfe,  1967 {\em ApJ} 147, 73 
\bibitem{Boughn & Crittenden 2004} S.~P.~Boughn \& R.~G.~Crittenden,  2004 {\em Nature} 427, 45
\bibitem{Boughn & Crittenden 2005} S.~P.~Boughn \& 
R.~G.~Crittenden,  2005 {\em NewAR} 49, 75 
\bibitem{Boldt 1987} E.~Boldt, 1987, {\em Phys. Rep.} 146, 215
\bibitem{Condon et al. 1998} J.~J.~Condon, et al. 1998 {\em AJ} 115, 1693 
\bibitem{Nolta et al. 2004}  M.~R.~Nolta, et al.,  2004 {\em ApJ} 608, 10\bibitem{Maddox et al. 1990} S.~J.~Maddox et al., 1990 {\em MNRAS}  242, 43 
\bibitem{York et al. 2000} D.~York et al. 2000 {\em AJ} 120, 1579
\bibitem{Jarrett et al. 2000} T.~H.~Jarrett et al. 2000 {\em ApJ} 119, 2498\bibitem{Fosalba et al. 2003} P.~Fosalba, , E.~Gazta{\~n}aga, \& F.~J.~Castander,  {\em ApJ}  597, L89
\bibitem{Scranton et al. 2003} R.~Scranton, et al. 2003 arXiv:astro-ph/0307335
\bibitem{Fosalba & Gaztanaga 2004} P.~Fosalba \& E.~Gazta{\~n}aga,  2004, {\em MNRAS} 350, L37
\bibitem{Afshordi et al. 2004} N.~Afshordi, Y.-S.~Loh, \& M.~A.~Strauss,  2004 {\em PhRvD} 69, 083524
\bibitem{Padmanabhan et al. 2005} N.~Padmanabhan et al., 2005 {\em PhRvD} 72, 043525
\bibitem{Cabre et al. 2006}  A.~Cabre, et al. 2006, arXiv:astro-ph/0603690 
\bibitem{Vielva et al. 2006} P.~Vielva, E.~Mart{\'{\i}}nez-Gonz{\'a}lez, 
\& M.~Tucci,  2006, {\em MNRAS} 365, 891
\bibitem{Kinney et al. 2006} W.~H. Kinney, E.~W. Kolb, A. Melchiorri, A. Riotto 2006, astro-ph/0605338
\bibitem{Seljak et al. 2006} U.~Seljak, A.~Slosar \& P.~McDonald 2006, astro-ph/0604335
\bibitem{Choudhury & Ferrara 2006} T.~Roy Choudhury \& A.~Ferrara 2006, astro-ph/0603149
\bibitem{Hu & Holder 2003} W.~Hu \& G.~P.~Holder, 2003, {\em Phys. Rev. D}, 68, 023001 
\bibitem{open questions} Peebles P.~J.~E., 2003, astro-ph/0311435
\end{thebibliography}
\end{document}